\documentclass[twoside]{dis07}
\usepackage[latin1]{inputenc}
\usepackage[dvips]{graphicx,epsfig,color}
\usepackage{wrapfig,rotating}
\usepackage{amssymb,amsmath,array}

\begin{document}
\title{Polarized Parton Densities and Higher Twist in the Light of
the Recent CLAS and COMPASS data }

\author{E. Leader$^1$, A. V. Sidorov$^2$ and D. B. Stamenov$^3$
%
\thanks{This work was supported by a UK Royal Society Joint
International Project Grant and RFBR Grants (No 05-01-00992,
06-02-16215, 07-02-01046 ).}
%
\vspace{.3cm}\\
%
1- Imperial College,
Prince Consort Road, London SW7 2BW, England \\
%
\vspace{.1cm}\\
2- Bogoliubov Theoretical Laboratory, Joint Institute for Nuclear
Research, 141980 Dubna, Russia\\
\vspace{.1cm}\\
3- Institute for Nuclear Research and Nuclear Energy, Bulgarian Academy of Sciences \\
Blvd. Tsarigradsko Chaussee 72, Sofia 1784, Bulgaria }


\maketitle

\begin{abstract}
The impact of the recent very precise CLAS and COMPASS $g_1/F_1$
data on polarized parton densities and higher twist effects is
discussed. We demonstrate that the low $Q^2$ CLAS data improve
essentially our knowledge of higher twist corrections to the spin
structure function $g_1$, while the large $Q^2$ COMPASS data
influence mainly the strange quark and gluon polarizations which
slightly decrease. We find also that the present inclusive DIS
data cannot rule out a negative polarized and changing in sign
gluon densities.
\end{abstract}

\section{Introduction}

One of the features of polarized DIS is that a lot of the present
data are in the preasymptotic region ($Q^2 \sim 1-5~\rm
GeV^2,~4~\rm GeV^2 < W^2 < 10~\rm GeV^2$). This is especially the
case for the experiments performed at the Jefferson Laboratory. As
was shown in \cite{LSS_HT}, to confront correctly the QCD
predictions to the experimental data including the preasymptotic
region, the {\it non-perturbative} higher twist (powers in
$1/Q^2$) corrections to the nucleon spin structure functions have
to be taken into account too.

In this talk we discuss the impact of the recent very precise CLAS
\cite{CLAS06} and COMPASS \cite{COMPASS06} inclusive polarized DIS
data on the determination of both the longitudinal polarized
parton densities (PDFs) in the nucleon and the higher twist (HT)
effects. These experiments give important information about the
nucleon structure in quite different kinematic regions. While the
CLAS data entirely belong to the preasymptotic region and as one
can expect they should mainly influence the higher twist effects,
the COMPASS data on the spin asymmetry $A_1^d$ are large $Q^2$
data and they should affect mainly the polarized parton densities.
In addition, due to COMPASS measurements we have for the first
time accurate data at small $x~(0.004 < x < 0.015)$, where the
behaviour of the spin structure function $g_1^d$ should be more
sensitive to the sign of the gluon polarization.

\section{Results of analysis}

The method used to extract simultaneously the polarized parton
densities and higher twist corrections to $g_1$ is described in
\cite{LSS_HT}. According to this method, the $g_1/F_1$ and
$A_1(\approx g_1/F_1)$ data have been fitted using the
experimental data for the unpolarized structure function
$F_1(x,Q^2)$
\begin{equation}
\left[{g_1(x,Q^2)\over F_1(x, Q^2)}\right]_{exp}~\Leftrightarrow~
{{g_1(x,Q^2)_{\rm LT}+h(x)/Q^2}\over F_1(x,Q^2)_{exp}}~.
\label{g1F2Rht}
\end{equation}
As usual, $F_1$ is replaced by its expression in terms of the
usually extracted from unpolarized DIS experiments $F_2$ and $R$
and the phenomenological parametrizations of the experimental data
for $F_2(x,Q^2)$ \cite{NMC} and the ratio $R(x,Q^2)$ of the
longitudinal to transverse $\gamma N$ cross-sections \cite{R1998}
are used. Note that such a procedure is equivalent to a fit to
$(g_1)_{exp}$, but it is more precise than the fit to the $g_1$
data themselves actually presented by the experimental groups
because here the $g_1$ data are extracted in the same way for all
of the data sets. In Eq. (\ref{g1F2Rht}) "LT" denotes the leading
twist contribution to $g_1$ (logarithmic in $Q^2$ NLO pQCD
expression where the target mass corrections are taken into
account), while $h(x)/Q^2$ corresponds to the first term in the
$(\Lambda^2_{\rm QCD}/Q^2)^n$ expansion of higher twist effects.

Let us discuss now how inclusion of the CLAS EG1 proton and
deuteron $g_1/F_1$ data \cite{CLAS06} and the {\it new} COMPASS
data on $A_1^d$ \cite{COMPASS06} influence our previous results
\cite{LSS05} on polarized PDFs and higher twist obtained from the
NLO QCD fit to the world data ({\it see} the references in
\cite{LSS05}), before the CLAS and the latest COMPASS data were
available.

\subsection{Impact of CLAS data}

As the CLAS data are mainly low $Q^2$ data where the role of HT
becomes important, they should help to fix better the higher twist
effects. Indeed, due to the CLAS data, the determination of HT
corrections to the proton and neutron spin structure functions,
$h^p(x)$ and $h^n(x)$, is significantly improved in the CLAS $x$
region, compared to the values of HT obtained from our LSS'05
analysis \cite{LSS05} in which a NLO(${\rm \overline {MS}}$) QCD
approximation for $g_1(x,Q^2)_{\rm LT}$ was used. This effect is
illustrated in Fig. 1. One can conclude now that
\begin{figure}[bht]
\centerline{ \epsfxsize=2.0in\epsfbox{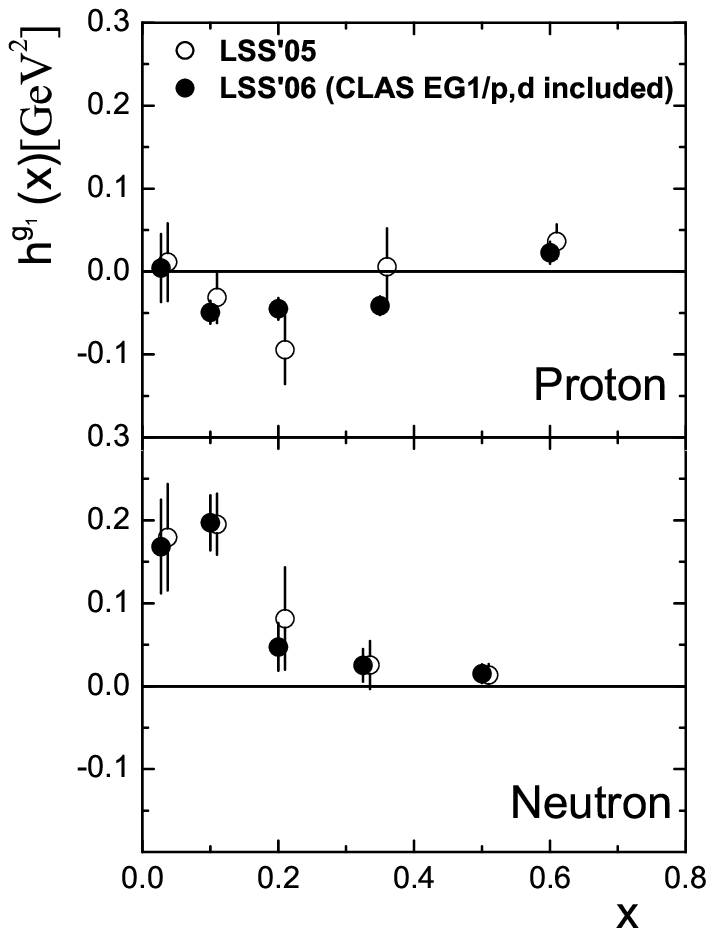}
\epsfxsize=2.0in\epsfbox{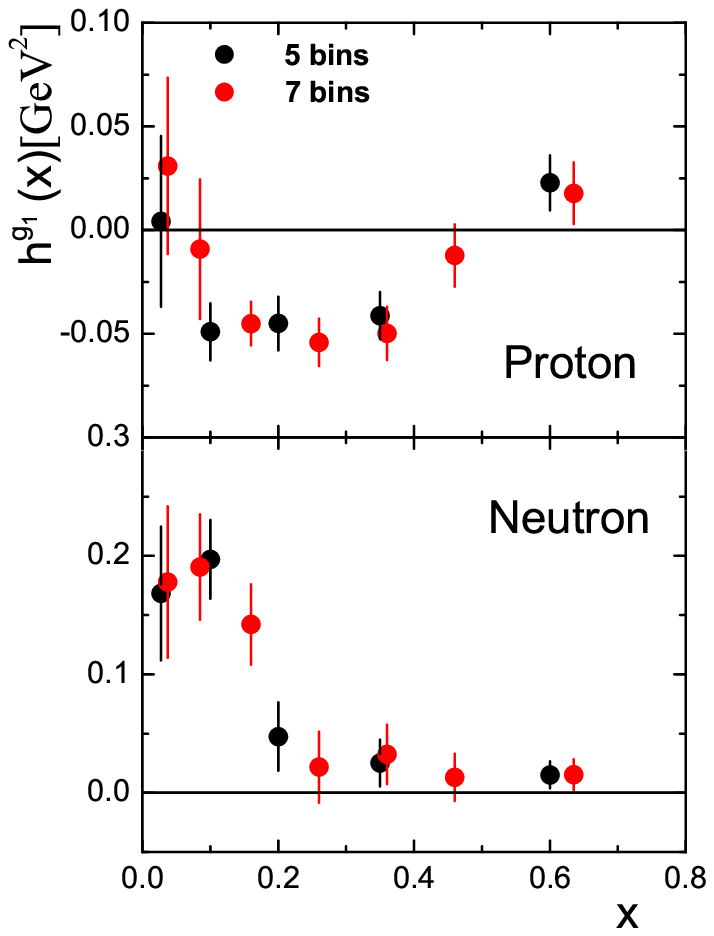} }
\caption{Effect of CLAS data on the higher twist values (left).
Comparison between the HT values corresponding to 5 and 7 bins
(right).}
\end{figure}
the HT corrections for the proton target are definitely different
from zero and negative in the $x$ region: 0.1-0.4. Also, including
the CLAS data in the analysis, the HT corrections for the neutron
target are better determined in the $x$ region: 0.2-0.4. Note that
$h^n(x)$ at $x \sim 0.5$ was already fixed very precisely from the
JLab Hall A data on the ratio $g^{(n)}_1/F^{(n)}_1$. As expected,
the central values of the polarized PDFs are practically {\it not}
affected by the CLAS data. This is a consequence of the fact that
at low $Q^2$ the deviation from logarithmic in $Q^2$ pQCD
behaviour of $g_1$ is accounted for by the higher twist term in
$g_1$. However, the accuracy of the determination of polarized PD
is essentially improved. This improvement (see \cite{url, LSS06})
is a consequence of the much better determination of higher twist
contributions to the spin structure function $g_1$, as discussed
above. Because of the good accuracy of the CLAS data, one can
split the measured $x$ region of the world+CLAS data set into 7
bins instead of 5, as used up to now, and therefore, can determine
more precisely the $x$-dependence of the HT corrections to $g_1$.
As seen in Fig. 1, the more detailed $x$-space behaviour of the HT
contribution, obtained when using 7 $x$-bins, suggests a smoother
function dependence in $x$ and will help us to calculate more
precisely their first moments in the experimental $x$ region and
to compare them with the predictions given by different
theoretical models.

The main message from this analysis is: It is impossible to
describe the very precise CLAS data if the HT corrections are NOT
taken into account. Note that if the low $Q^2$ data are not too
accurate, it would be possible to describe them using only the
leading twist term (logarithmic in $Q^2$) in $g_1$, {\it i.e.} to
mimic the power in $Q^2$ dependence of $g_1$ with a logarithmic
one (using different forms for the input PDFs and/or more free
parameters associated with them) which was done in the analyses of
another groups before the CLAS data have appeared.

\subsection{Impact of new COMPASS data}

In contrast to the CLAS data, the COMPASS data are mainly at large
$Q^2$ and the only precise data at small $x: 0.004 < x < 0.02$.
The new data are based on 2.5 times larger statistics than those
of COMPASS'05 and give more precise and detailed information about
$A^d_1$ at small $x$ (see Fig. 2a).
\begin{figure}[bht]
\centerline{ \epsfxsize=2.3in\epsfbox{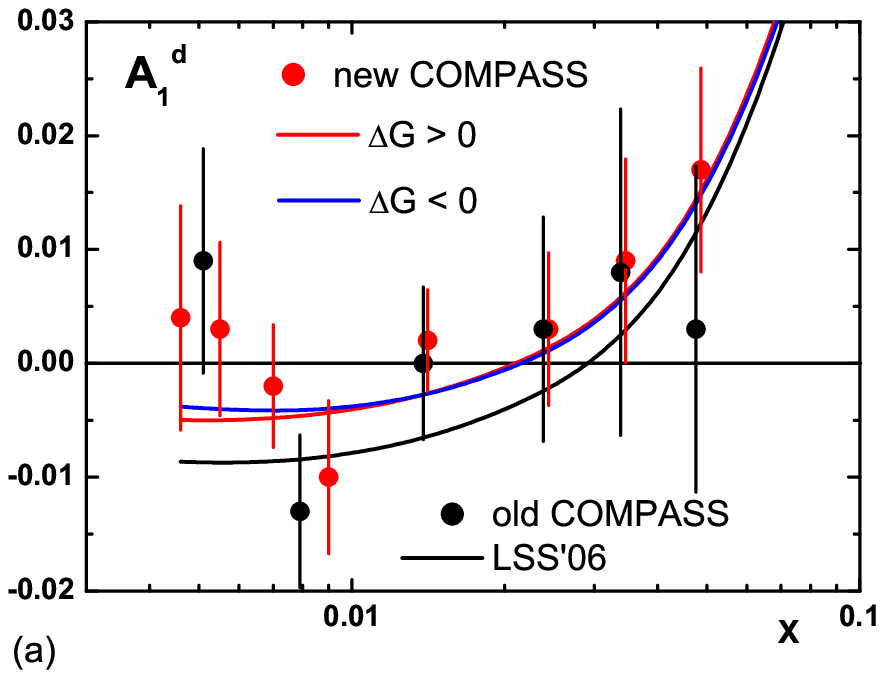}
\epsfxsize=2.2in\epsfbox{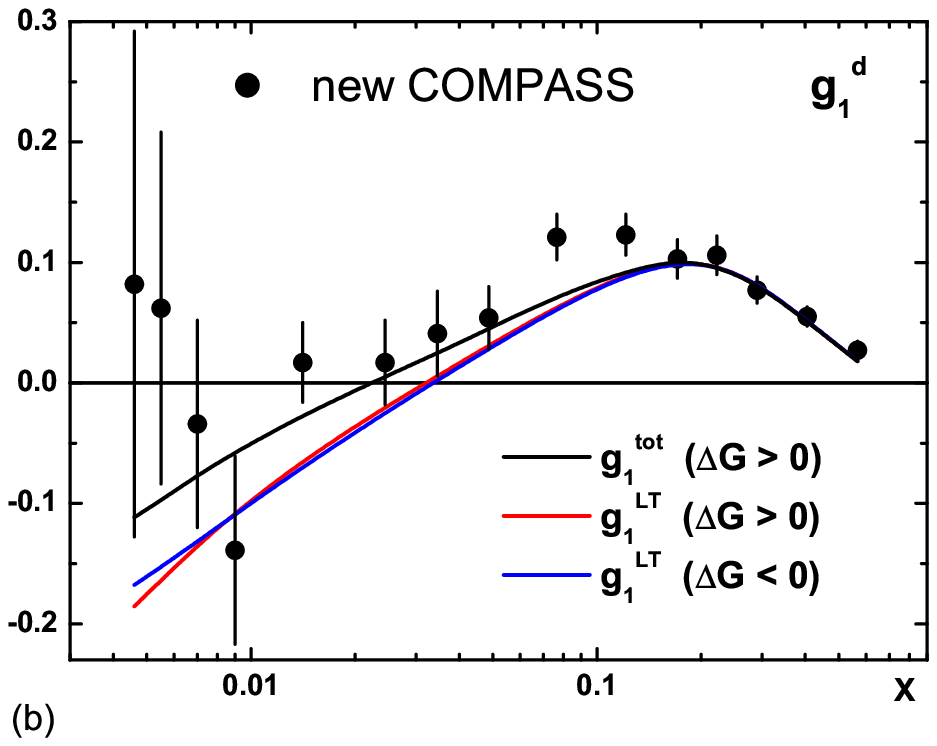} }
 \caption{Comparison of our NLO($\rm \overline{MS}$) results for
 $A^d_1$ (a) and $g^d_1$ (b) corresponding to $\Delta G > 0$
 and $\Delta G < 0$ with the new COMPASS data at measured $x$ and
 $Q^2$ values. }
\end{figure}
The QCD theoretical curves for $A^d_1$ corresponding to the best
fits with positive and negative $\Delta G$ lie above the old one
at $x < 0.1$. As a result, the COMPASS'06 data do not influence
$(\Delta u + \Delta\bar u)$ and $(\Delta d + \Delta\bar d)$ parton
densities, while the magnitudes of both the polarized gluon and
strange quark sea densities and their first moments slightly
decrease (see Refs. \cite{url, LSS06}). As a consequence, $\Delta
\Sigma(Q^2= 1~\rm GeV^2)$ increases from (0.165 $\pm$ 0.044) to
(0.207 $\pm$ 0.040) for $\Delta G >0$ and (0.243 $\pm$ 0.065) for
$\Delta G < 0$.
\begin{wrapfigure}{r}{0.5\columnwidth}
\centerline{\includegraphics[width=0.45\columnwidth]{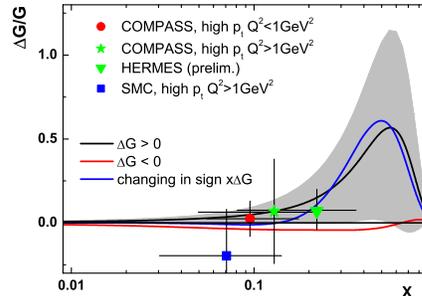}}
\caption{Comparison between the experimental data and
 NLO($\rm \overline{MS}$) curves
 for the gluon polarization $\Delta G(x)/G(x)$ at $Q^2=3~\rm GeV^2$
 corresponding to $\Delta G > 0$, $\Delta G < 0$ and an
 oscillating-in-sign $x\Delta G$. Error bars
 represent the total (statistical and systematic) errors. }
\label{delGovG}
\end{wrapfigure}
As expected, the values of HT are practically not affected by
COMPASS data excepting the small $x$ where $Q^2$ are also small.
We have found that the HT effects at small $x$ are large, up to
40\% of the magnitude of $(g^d_1)_{\rm LT}$ (see Fig. 2b) and
therefore, in the presence of HT, $~g^d_1$, as well as $~A^d_1$
(Fig. 1a), are not too sensitive to the sign of the gluon
polarization at small $x$. These results are in contrast to those
obtained in the COMPASS analysis \cite{COMPASS06} where the higher
twist corrections are not taken into account (for details see
\cite{url, LSS06}).

We have observed that the present inclusive DIS data cannot rule
out the solutions with negative and changing in sign gluon
polarizations. The shape of the negative gluon density differs
from that of positive one. In all the cases the magnitude of
$\Delta G$ is small: $|\Delta G| \leq  0.2$ and the corresponding
polarized quark densities are very close to each other. In Fig. 3
the ratio $\Delta G(x)/G(x)$ calculated for the different $\Delta
G(x)$ obtained in our analysis and using $G(x)_{\rm MRST'02}$ for
the unpolarized gluon density, is compared to the existing direct
measurements of $\Delta G/G$. The most precise value for $\Delta
G/G$, the COMPASS one, is well consistent with any of the
polarized gluon densities determined in our analysis.

{\bf In conclusion}, it was demonstrated that the inclusion of the
low $Q^2$ CLAS data in the NLO QCD analysis of the world DIS data
improves essentially our knowledge of HT corrections to $g_1$ and
does not affect the central values of PDFs, while the large $Q^2$
COMPASS data influence mainly the strange quark and gluon
polarizations, but practically do not change the HT corrections.
These results strongly support the QCD framework, in which the
leading twist pQCD contribution is supplemented by higher twist
terms of ${\cal O}(\Lambda^2_{\rm QCD}/Q^2)$.

\begin{footnotesize}

\end{footnotesize}

\end{document}